\documentclass[preprint,prd,aps,showpacs,showkeys,nofootinbib]{revtex4}
\usepackage{graphicx}
\begin{document}


\title{Muon $(g-2)$ in the B-LSSM}

\author{Jin-Lei Yang$^{1,2}$\footnote{yangjinlei@itp.ac.cn}, Hai-Bin Zhang$^{3,4}$\footnote{hbzhang@hbu.edu.cn}, Chang-Xin Liu$^{3,4}$, Xing-Xing Dong$^{3,4}$,
Tai-Fu Feng$^{3,4,5}$\footnote{fengtf@hbu.edu.cn}}

\affiliation{CAS Key Laboratory of Theoretical Physics, Institute of Theoretical Physics, Chinese Academy of Sciences, Beijing 100190, China$^1$\\
School of Physical Sciences, University of Chinese Academy of Sciences, Beijing 100049, China$^2$\\
Department of Physics, Hebei University, Baoding, 071002, China$^3$\\
Key Laboratory of High-precision Computation and Application of Quantum Field Theory of Hebei Province, Baoding, 071002, China$^4$\\
Department of Physics, Chongqing University, Chongqing 401331, China$^5$}

\begin{abstract}
The difference between the updated experimental result on the muon anomalous magnetic dipole moment and the corresponding theoretical prediction of the standard model on that is about $4.2$ standard deviations. In this work, we calculate the muon anomalous MDM at the two-loop level in the supersymmetric $B-L$ extension of the standard model. Considering the experimental constraints on the lightest Higgs boson mass, Higgs boson decay modes $h\rightarrow \gamma\gamma,\;WW,\;ZZ,\; b\bar b,\;\tau\bar\tau$, B rare decay $\bar B\rightarrow X_s\gamma$, and the transition magnetic moments of Majorana neutrinos, we analyze the theoretical predictions of the muon anomalous magnetic dipole moment in the $B-L$ supersymmetric model. The numerical analyses indicate that the tension between the experimental measurement and the standard model prediction is remedied in the $B-L$ supersymmetric model.

\end{abstract}


\maketitle

\section{Introduction\label{sec1}}
\indent\indent
Recently, the Muon $g-2$ experiment at Fermilab~\cite{MDM-exp1,MDM-exp2,MDM-exp3,MDM-exp4} has measured the muon anomalous magnetic dipole moment (MDM), $a_\mu=(g_\mu-2)/2$, and the reported result based on its Run-1 data is $3.3$ standard deviations greater than the standard model (SM) prediction. The result agrees with the previous Brookhaven National Laboratory E821 measurement~\cite{muon-exp} very well. The new experimental average for the difference between the experimental measurement and SM theoretical prediction of the muon anomalous MDM is given by
\begin{eqnarray}
\Delta a_{\mu}=a_{\mu}^{\rm exp}-a_{\mu}^{\rm SM}=(25.1\pm5.9)\times10^{-10},
\end{eqnarray}
which shows that the tension between experiment and the SM prediction is increased to 4.2 standard deviations. Then many papers appeared to study the relation between the updated muon anomalous MDM results with various models beyond the SM, the details can be seen in Refs.~\cite{Jegerlehner:2009ry,Davier:2010nc,Eidelman:1995ny,Moroi:1995yh,Aoyama:2012wk,Hagiwara:2006jt,
Bijnens:1995xf,Martin:2001st,Hagiwara:2003da,Bijnens:2001cq,Chattopadhyay:1995ae,Davier:2017zfy,
Bijnens:1995cc,Hagiwara:2002ma,Hayakawa:1995ps,Hisano:2001qz,Everett:2001tq,deRafael:1993za,Chattopadhyay:2001vx,
Hayakawa:1996ki,DeTroconiz:2001rip,Baek:2001kca,Davoudiasl:2012ig,RamseyMusolf:2002cy,Benayoun:2012wc,
Colangelo:2014qya,Komine:2001fz,Kinoshita:2005sm,Wang:2015kuj,Allanach:2015gkd,Belanger:2001am,
Chang:2000ii,Aoyama:2007dv,Passera:2006gc,Cao:2019evo,Abdughani:2019wai,Wang:2018vrr,Wang:2017vxj,Kiritsis:2002aj,
Padley:2015uma,Li:2018aov,Li:2020dbg,Cao:2021lmj,Chen:2021rnl,Yin:2021yqy,Yin:2020afe,Sabatta:2019nfg,
vonBuddenbrock:2019ajh,vonBuddenbrock:2016rmr,Okada:2016wlm,Fukuyama:2016mqb,Belanger:2017vpq,Megias:2017dzd,Tran:2018kxv,
g-2muonQCD,g-2muon,g-2muon1,g-2muon2,g-2muon3,g-2muon4,g-2muon5,g-2muon6,g-2muon7,g-2muon8,g-2muon9,g-2muon10,
g-2muon11,g-2muon12,g-2muon13,g-2muon14,g-2muon15,g-2muon16,g-2muon17,g-2muon18,g-2muon19,g-2muon20,
g-2muon21,g-2muon22,g-2muon23,g-2muon24,g-2muon25,g-2muon26,g-2muon27,g-2muon28,g-2muon29,g-2muon30}. However, it is worth mentioning that the latest result obtained by the lattice QCD calculation~\cite{Borsanyi:2020mff} of the
leading order hadronic vacuum polarization contribution to the muon anomalous MDM is larger than the former result, which can accommodate the discrepancy between the SM prediction and the experimental result, hence the discrepancy needs further scrutiny.

It is well-known that the muon anomalous MDM has close relation with the new physics (NP) beyond the SM, and the tiny neutrino masses shown in neutrino oscillation experiments~\cite{nu2} are an unambiguous evidence of NP. The transition magnetic moment is one of the most important properties of massive Majorana neutrinos, which has significant astrophysical consequences even if the value of neutrino transition magnetic moment is extremely small. Hence, we focus on the NP contributions to the muon anomalous MDM and the transition magnetic moments of Majorana neutrinos in the supersymmetric extension of the SM with local $B-L$ gauge symmetry (B-LSSM).

The gauge group of the B-LSSM is $SU(3)_C\bigotimes SU(2)_L\bigotimes U(1)_Y\bigotimes U(1)_{B-L}$. It is obvious that new gauge group $U(1)_{B-L}$~\cite{Davidson:1978pm} is added in the B-LSSM, which introduces a new neutral gauge boson $Z'$~\cite{Chun:2018ibr}. In addition, there are two Higgs singlets and three right-handed neutrinos, then tiny neutrino masses can be obtained naturally in the B-LSSM by the so-called type-I see-saw mechanism. The local gauge group $U(1)_{B-L}$ is broken when new Higgs singlets $\eta_1,\;\eta_2$ receive vacuum expectation values (VEVs) $u_1,\;u_2$ (the definitions are similar to the two Higgs doublets VEVs $v_1,\;v_2$), and we can define $\tan\beta'=u_2/u_1$ (analogy to the definition of $\tan\beta=v_2/v_1$ in the minimal supersymmetric SM (MSSM)). The super partners of new introduced Higgs singlets and neutral gauge boson can be dark matter candidates~\cite{16,1616,DelleRose:2017ukx,DelleRose:2017uas}.

With respect to the MSSM, the B-LSSM has richer phenomenology due to the existence of new neutral gauge boson $Z'$, Majorana neutrinos, new dark matter candidates, and new Higgs bosons. The discovery potential of the model by the future runs of the  Large Hadron Collider (LHC) has been discussed in many works. Ref.~\cite{Basso:2008iv} shows the $Z'$ decay chain involving heavy neutrinos, and eventually decaying into leptons and jets allows one to measure the $Z'$ and heavy neutrino masses at the LHC. The direct production of right-handed sneutrinos at the LHC and their decay modes are studied in Ref.~\cite{Elsayed:2012ec}. The second light scalar Higgs signal in the decays of the SM-like Higgs to $\gamma\gamma$ and $Z\gamma$ at the CERN machine are discussed in Ref.~\cite{Hammad:2015eca}. The disentangling of the B-LSSM with other SUSY models at the CERN machine is analyzed in Ref.~\cite{Abdallah:2015uba}, which also shows that the mono-jet events can be accessible at the LHC. Ref.~\cite{Abdallah:2018kix} shows that the LHC will enable to establish a specific B-LSSM signal during Run $2$ and $3$, mediated by a charged Higgs boson pair produced from the on-shell $Z'$ decay. Considering the lightest supersymmetric particle (LSP) neutralino, Ref.~\cite{Ahmed:2020lua} shows that the LSP relic density constraint provides a lower bound on the stop and gluino masses of about $3$ TeV and $4$ TeV respectively, which is testable in the near future collider experiments such as high luminosity LHC.

The framework of the B-LSSM has been discussed detailedly in our previous works~\cite{Yang:2018utw,Yang:2019aao,Yang:2020ebs}, which contain the particle content, the superpotential, the soft breaking terms, some mass matrices and interaction vertices. We do not introduce the model in detail in this work, but some relevant mass matrices will be given in Sec.~\ref{sec2} for analyzing the numerical results clearly. Then we also give a brief discussion about the contributions to muon anomalous MDM and Majorana neutrinos transition magnetic moments in the B-LSSM in Sec.~\ref{sec2}. Considering the constraints from the measured Higgs boson mass and etc, we present the numerical results of the NP contributions to the muon anomalous MDM and the Majorana neutrinos transition magnetic moments in the B-LSSM in Sec.~\ref{sec3}. Finally, a brief summary is given in Sec.~\ref{sec4}. Some mass matrices are collected in the appendix.

\section{$\bigtriangleup a_\mu^{NP}$ and transition magnetic moment of Majorana neutrinos in the B-LSSM\label{sec2}}

\begin{figure}
\setlength{\unitlength}{1mm}
\centering
\includegraphics[width=6in]{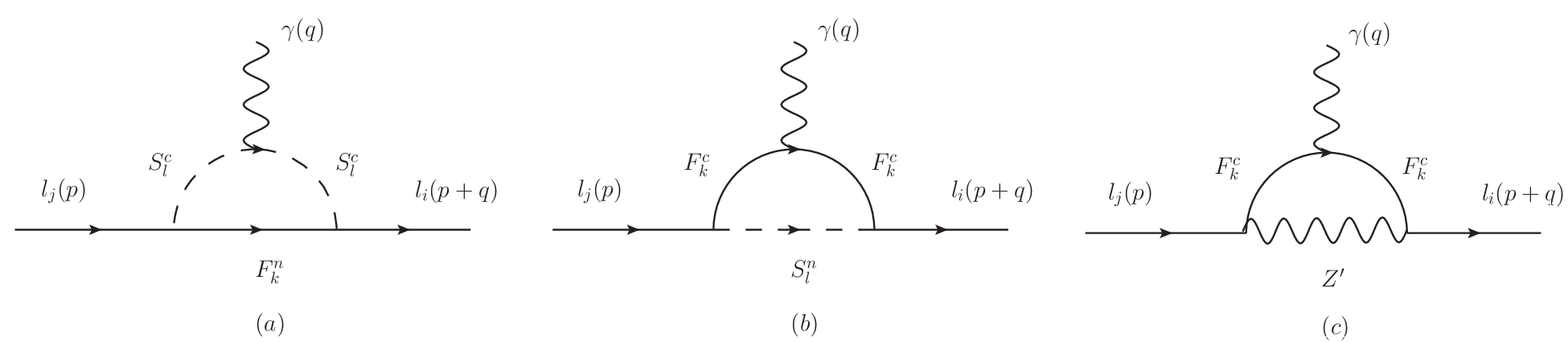}
\vspace{0cm}
\caption[]{One-loop Feynman diagrams which contribute to the lepton MDM. (a) represents the contributions to $\bigtriangleup a_l^{NP}$ from neutral fermions, (b) represents the contributions from charged fermions, (c) represents the contributions from new $Z'$ boson in the B-LSSM.}
\label{one loop Feynman diagram}
\end{figure}

\begin{figure}
\setlength{\unitlength}{1mm}
\centering
\includegraphics[width=6in]{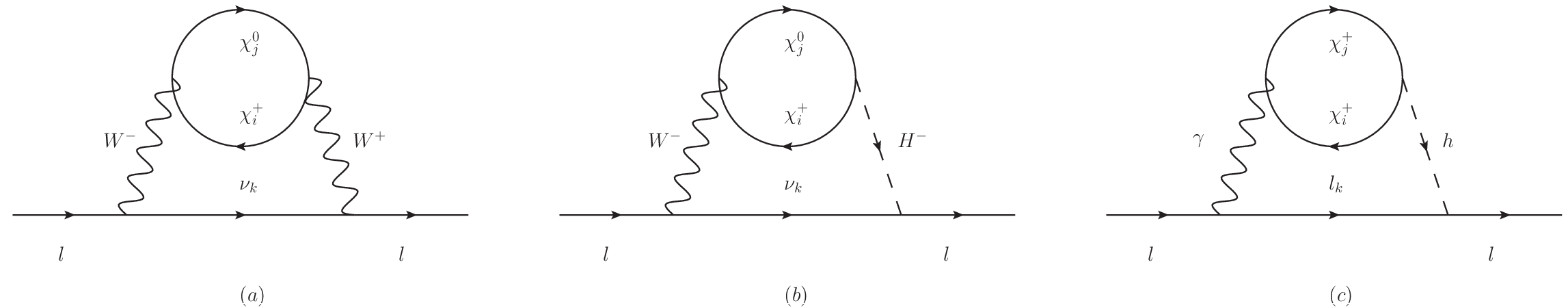}
\vspace{0cm}
\caption[]{The two-loop Barr-Zee type diagrams which contribute to the lepton MDM, the corresponding contributions to $\bigtriangleup a_l^{NP}$ are obtained by attaching a photon to the internal particles in all possible ways.}
\label{two loop Feynman diagram}
\end{figure}

The one-loop contributions to the muon anomalous MDM in the B-LSSM are depicted in Fig.~\ref{one loop Feynman diagram}.
At the one-loop level, the dominant contributions corresponding to Fig.~\ref{one loop Feynman diagram} (a) come from the neutralino-charged slepton loop, the dominant contributions corresponding to Fig.~\ref{one loop Feynman diagram} (b) come from the chargino-sneutrino loop, and Fig.~\ref{one loop Feynman diagram} (c) represents the contributions from new $Z'$ boson in the B-LSSM. In addition, the numerical results of our previous work~\cite{Yang:2018guw} show that the contributions from the two-loop Barr-Zee type diagrams are important to the muon anomalous MDM. Hence we also consider the two-loop corrections to $\Delta a_\mu^{NP}$, and the corresponding Feynman diagrams are shown in Fig.~\ref{two loop Feynman diagram}.
Then the NP contributions to the muon anomalous MDM in the B-LSSM can be written as
\begin{eqnarray}
\Delta a_{\mu}^{NP}=\Delta a_{\mu}^{\rm one-loop}+\Delta a_{\mu}^{\rm two-loop},
\end{eqnarray}
where the concrete expressions of $\Delta a_{\mu}^{\rm one-loop},\;\Delta a_{\mu}^{\rm two-loop}$ can be found in our previous works~\cite{Yang:2018guw,hLFV,hZr}.

\begin{figure}
\setlength{\unitlength}{1mm}
\centering
\includegraphics[width=4.5in]{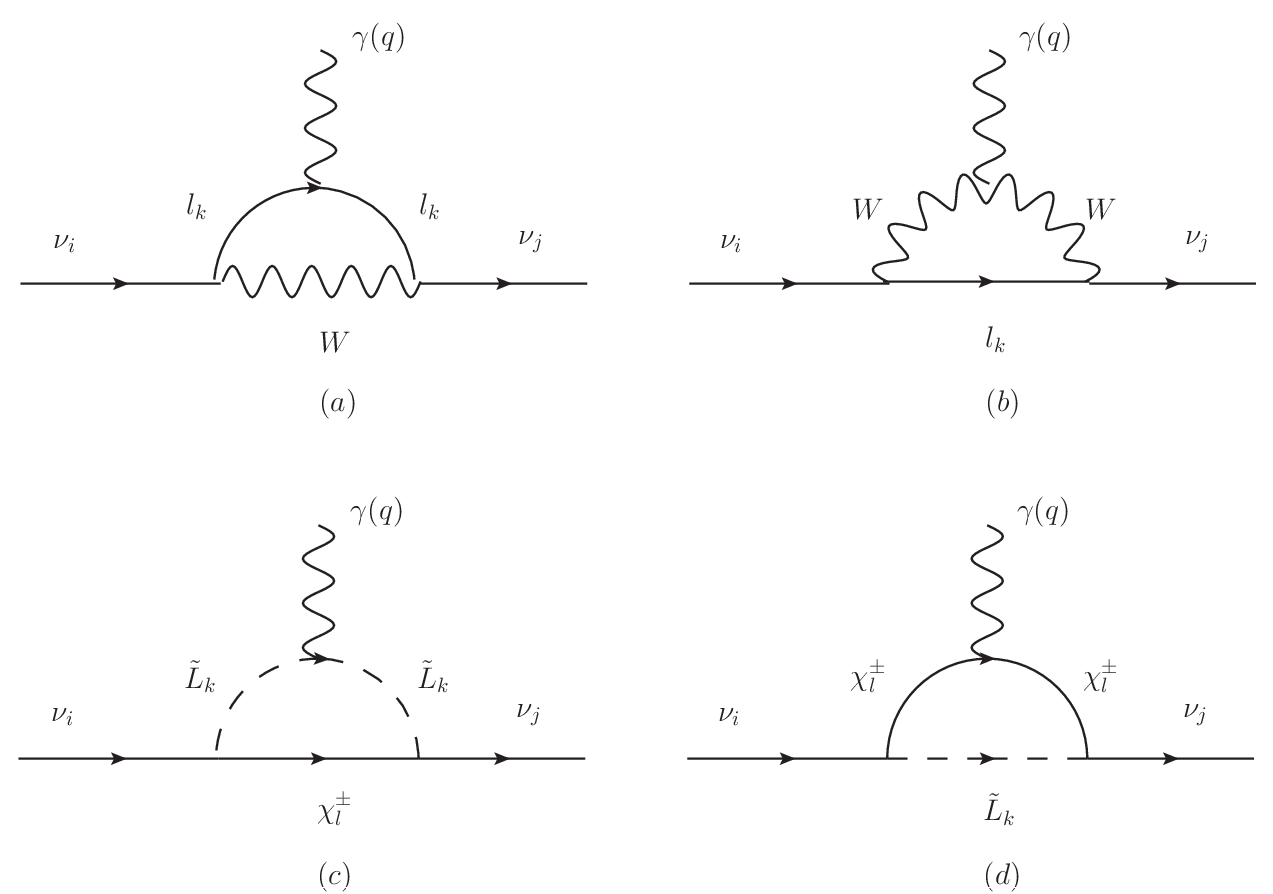}
\vspace{0cm}
\caption[]{The one-loop diagrams which contribute the Majorana neutrinos transition magnetic moments, where $\tilde L$ denotes charged slepton, $\chi^\pm$ denotes chargino, $l$ denotes charged lepton.}
\label{nuinujr}
\end{figure}

For the transition magnetic moments of Majorana neutrinos, the contributions in the B-LSSM can be written as
\begin{eqnarray}
\mu_{ij}^M=\mu_{ij}^D-\mu_{ji}^D,
\end{eqnarray}
with
\begin{eqnarray}
\mu_{ij}^D=4m_e m_{\nu i}\Re(\frac{m_{\nu j}}{m_{\nu i}}C_2^{L*})\mu_B,
\end{eqnarray}
where $i,j$ are the indices of generation, $m_e$ is the electron mass, $m_{\nu i}$ is the light neutrino mass, $\mu_B\equiv e/(2m_e)$, and $C_2^{L}$ is the coefficient of operator ${\mathcal O}_2^{L}$
\begin{eqnarray}
{\mathcal O}_2^{L}=e\overline{(i{\mathcal D}_\mu \nu_i)}\gamma^\mu F\cdot\sigma P_L\nu_j,
\end{eqnarray}
where ${\mathcal D}_\mu=\partial_\mu+ieA_\mu$, $P_L=(1-\gamma_5)/2$, $\sigma^{\mu\nu}=\frac{i}{2}[\gamma^\mu,\gamma^\nu]$, $F_{\mu\nu}$ is the electromagnetic field strength, $\nu_{i,j}$ is the four component neutrinos. The Feynman diagrams which contribute to the transition magnetic moments of Majorana neutrinos in the B-LSSM are plotted in Fig.~\ref{nuinujr}, then the coefficient $C_2^{L}$ can be written correspondingly as
\begin{eqnarray}
C_2^L=C_2^{L(a)}+C_2^{L(b)}+C_2^{L(c)}+C_2^{L(d)},
\end{eqnarray}
where the concrete expressions of $C_2^{L(a,...,d)}$ can be found in our previous work~\cite{Zhang:2014iva}.

Refs.~\cite{Yang:2018guw,hLFV,hZr,Zhang:2014iva,Yang:2020bmh,Dong:2020ioc} show that the theoretical predictions of the muon anomalous MDM and the transition magnetic moments of Majorana neutrinos in the B-LSSM depend on sleptons, neutralinos and charginos strongly. On the basis $(\tilde e_L,\;\tilde e_R)$, the mass matrix of charged sleptons is given as:
\begin{eqnarray}
&&m_{\tilde e}^2=
\left(\begin{array}{cc}m_{eL}^2,&\frac{1}{\sqrt2}(v_1 T_e^\dagger-v_2\mu Y_e^\dagger)\\\frac{1}{\sqrt2}(v_1 T_e-v_2\mu^* Y_e),&m_{eR}^2\end{array}\right),\label{ME1}
\end{eqnarray}
\begin{eqnarray}
&&m_{eL}^2=\frac{1}{8}\Big[2g_{_B}(g_{_B}+g_{_{YB}})(u_1^2-u_2^2)+(g_1^2-g_2^2+g_{_{YB}}^2+2g_{_B}g_{_{YB}})(v_1^2-
v_2^2)\Big]\nonumber\\
&&\qquad\;\quad\;+m_{\tilde L}^2+\frac{v_1^2}{2}Y_e^\dagger Y_e,\nonumber\\
&&m_{eR}^2=\frac{1}{24}\Big[2g_{_B}(g_{_B}+2g_{_{YB}})(u_2^2-u_1^2)+2(g_1^2+g_{_{YB}}^2+2g_{_B}g_{_{YB}})(v_2^2-
v_1^2)\Big]\nonumber\\
&&\qquad\;\quad\;+m_{\tilde e}^2+\frac{v_1^2}{2}Y_e^\dagger Y_e,\label{ME2}
\end{eqnarray}
where $g_{_B}$ is the $U(1)_{B-L}$ coupling constant, $g_{_{YB}}$ is the gauge kinetic mixing coupling constant which arises from the existence of two Abelian gauge groups, $u_{1,2},\;v_{1,2}$ are the VEVs of two Higgs singlets and two Higgs doublets respectively (the definitions analogy to ones in the MSSM), and $T_e$ is the trilinear Higgs slepton coupling. In the following analyses, we assume the degenerate slepton masses and take $m_{\tilde L}=m_{\tilde e}=m_{\tilde \nu}={\rm diag}(M_E,M_E,M_E)\;{\rm TeV}$ for simplicity, where $m_{\tilde \nu}$ is the sneutrino mass term in sneutrino mass matrices as shown in the appendix.

For the mass matrix of charginos, the mass term in the B-LSSM can be written as
\begin{eqnarray}
&&{\mathcal L}_{M_{\chi^\pm}}=M_2 \tilde W^- \tilde W^++\mu \tilde H_1^- \tilde H_2^++\frac{1}{\sqrt 2}g_2 v_2\tilde W^-\tilde H_2^++\frac{1}{\sqrt 2}g_2 v_1\tilde W^+\tilde H_1^-+h.c.\nonumber\\
&&\qquad\quad=\left(\begin{array}{cc}\tilde W^-,&\tilde H_1^-\end{array}\right)\left(\begin{array}{cc}M_2,&\frac{1}{\sqrt 2}g_2 v_2\\ \frac{1}{\sqrt 2}g_2 v_1,&\mu\end{array}\right)\left(\begin{array}{cc}\tilde W^+\\ \tilde H_2^+\end{array}\right)+h.c.,
\end{eqnarray}
Four $2-$component spinors $(\tilde W^-,\tilde W^+,\tilde H_1^-,\tilde H_2^+)$ combine to give two $4-$component interaction eigenstate $\chi_1^I,\chi_2^I$. Then the mass term can be rewritten as
\begin{eqnarray}
&&{\mathcal L}_{M_{\chi^\pm}}=\left(\begin{array}{cc} {\bar\chi}_1^I,&{\bar\chi}_2^I\end{array}\right)\left(\begin{array}{cc}M_2,&\frac{1}{\sqrt 2}g_2 v_2\\ \frac{1}{\sqrt 2}g_2 v_1,&\mu\end{array}\right)\left(\begin{array}{cc}\chi_1^I\\ \chi_2^I\end{array}\right),
\end{eqnarray}
where $\bar\chi_i^I=\chi_i^{I\dagger}\gamma^0$, $\chi_1^I=\left(\begin{array}{cc}\tilde W^+\\ \overline{\tilde W}^-\end{array}\right)$ and $\chi_2^I=\left(\begin{array}{cc}\tilde H_2^+\\ \overline{\tilde H}_1^-\end{array}\right)$. Then the physical masses of $\chi_1,\chi_2$ ($\chi_{1},\chi_{2}$ are defined as mass eigenstates) can be obtained by diagonalizing the mass matrix
\begin{eqnarray}
&&M_{\chi^\pm}=
\left(\begin{array}{cc}M_2,&\frac{1}{\sqrt 2}g_2 v_2\\\frac{1}{\sqrt 2}g_2 v_1,&\mu\end{array}\right).
\end{eqnarray}
On the basis $(\tilde B,\;\tilde W^0,\;\tilde H_1^0,\;\tilde H_2^0,\;\tilde B',\;\tilde \eta_1,\;\tilde \eta_2)$, the mass matrix of neutralinos reads
\begin{eqnarray}
&&M_{\chi^0}=
\left(\begin{array}{ccccccc}M_1,&0,&-\frac{1}{2}g_1v_1,&\frac{1}{2}g_1v_2,&M_{BB'},&0,&0\\
0,&M_2,&\frac{1}{2}g_2v_1,&-\frac{1}{2}g_2v_2,&0,&0,&0\\
-\frac{1}{2}g_1v_1,&\frac{1}{2}g_2v_1,&0,&-\mu,&-\frac{1}{2}g_{_{YB}}v_1,&0,&0\\
\frac{1}{2}g_1v_2,&-\frac{1}{2}g_2v_2,&-\mu,&0,&\frac{1}{2}g_{_{YB}}v_2,&0,&0\\
M_{BB'},&0,&-\frac{1}{2}g_{_{YB}}v_1,&\frac{1}{2}g_{_{YB}}v_2,&M_{BL},&-g_{_B} u_1,&g_{_B} u_2\\
0,&0,&0,&0,&-g_{_B} u_1,&0,&-\mu'\\
0,&0,&0,&0,&g_{_B} u_2,&-\mu',&0\end{array}\right).
\end{eqnarray}
For gaugino mass terms $M_1,\;M_2,\;\mu$ appeared in the mass matrices of chargino and neutralino, we take $m_0\equiv \mu=2M_1=2M_2$ in the following analyses for simplicity. It can be noted that $m_0$ plays an important role in the contributions to the muon anomalous MDM through both of the chargino-sneutrino loop and the neutralino-charged slepton loop.

In the numerical calculation, we also consider the constraints from SM-like Higgs boson mass, Higgs boson decay modes $h\rightarrow \gamma\gamma,\;WW,\;ZZ,\;b\bar b,\;\tau\bar\tau$, and the B meson rare decay $\bar B\rightarrow X_s\gamma$. The stop and sbottom quark affect the Higgs boson mass obviously in supersymmetric models, and the $B$ meson rare decay $\bar B\rightarrow X_s\gamma$ also depends on the up-squark sector in the B-LSSM as shown in Ref.~\cite{Yang:2018fvw}. In order to explain the considered constraints clearly, the mass matrices of up-squark and down-squark in the B-LSSM are given in appendix~\ref{appA}. In addition, the leading-log radiative corrections from stop, top quark and sbottom, bottom quark to the mass of the SM-like Higgs boson are considered~\cite{70,71,72}.

\section{Numerical analyses\label{sec3}}
In the calculation, we take the $W$ boson mass $m_W=80.385\;{\rm GeV}$, the $Z$ boson mass $m_Z=90.19\;{\rm GeV}$, the bottom quark mass $m_b=4.65\;{\rm GeV}$, the top quark mass $m_t=173.5\;{\rm GeV}$, the electron mass $m_e=0.511\;{\rm MeV}$, the muon mass $m_\mu=0.106\;{\rm GeV}$, $\alpha_{em}(m_Z)=1/128.9$, $\alpha_s(m_Z)=0.118$. The constraint on the sum of neutrino masses $\sum_i m_{\nu i}<0.15\;{\rm eV}$~\cite{Ade:2015xua} is considered, and the neutrino mass-squared differences at $3\sigma$ level errors read~\cite{Esteban:2018azc}
\begin{eqnarray}
&&\Delta m_{12}^2=m_{\nu 2}^2-m_{\nu 1}^2=(7.4\pm0.61)\times 10^{-5}\;{\rm eV}^2,\nonumber\\
&&|\Delta m_{23}^2|=|m_{\nu 3}^2-m_{\nu 2}^2|=(2.52\pm0.1)\times 10^{-3}\;{\rm eV}^2.
\end{eqnarray}
Since the hierarchy of neutrino masses has not been fixed yet, we will take $m_{\nu1}<m_{\nu2}<m_{\nu3}$ for the normal hierarchy (NH) and $m_{\nu3}<m_{\nu1}<m_{\nu2}$ for the inverse hierarchy (IH) in the following analyses. The light neutrino mixing matrix is taken as the Pontecorvo-Maki-Nakagawa-Sakata mixing matrix.

In addition, the measured Higgs boson mass at $3\sigma$ level errors reads
\begin{eqnarray}
&&m_h=125.09\pm0.72\;{\rm GeV}.
\end{eqnarray}
For the signal strengths of the lightest Higgs boson decay modes $h\rightarrow \gamma\gamma,\;WW,\;ZZ,\; b\bar b,\;\tau\bar\tau$, we adopt the averages of the results from PDG which reads~\cite{53,54,55,56,57,58,59,60,61}
\begin{eqnarray}
&&\mu_{\gamma\gamma}^{exp}=1.11_{-0.09}^{+0.10},\;\mu_{WW}^{exp}=1.19\pm0.12,\;
\mu_{ZZ}^{exp}=1.20_{-0.11}^{+0.12},\nonumber\\
&&\mu_{b\bar b}^{exp}=1.04\pm0.13,\;\mu_{\tau\bar\tau}^{exp}=1.15_{-0.15}^{+0.16}.\label{hdecay}
\end{eqnarray}
The current combined experimental data for the branching ratio of $\bar B\rightarrow X_s\gamma$ reads~\cite{PDG1}
\begin{eqnarray}
&&{\rm Br}(\bar B\rightarrow X_s\gamma)=(3.49\pm0.19)\times10^{-4}.
\end{eqnarray}

In the previous section, we assume the slepton mass parameters as $m_{\tilde L}=m_{\tilde e}=m_{\tilde \nu}={\rm diag}(M_E,M_E,M_E)\;{\rm TeV}$ and the gaugino mass parameters as $m_0\equiv\mu=2M_1=2M_2$. There are constraints on the chargino and slepton masses from the LHC~\cite{Aad:2020qnn,Sirunyan:2020eab,Aad:2021ajl}. The results of CMS search~\cite{Sirunyan:2020eab} allow the exclusion of chargino $\tilde\chi_1^\pm$ (neutralino $\tilde\chi_2^0$) mass up to $750\;(800)\;{\rm GeV}$ via the on-shell decay to lightest neutralino $\tilde\chi_1^0$ and $W\;(Z)$ boson. And more recently, the CMS search~\cite{Aad:2021ajl} obtains a lower bound of $m_{\tilde\chi_1^\pm/\tilde\chi_2^0}>300\;{\rm GeV}$ via the off-shell decay of $\tilde\chi_1^\pm/\tilde\chi_2^0$ to $\tilde\chi_1^0$ and $W/Z$. Under our assumption, i.e. $m_0\equiv\mu=2M_1=2M_2$, we have $m_{\tilde\chi_1^\pm}\approx m_{\tilde\chi_2^0}\approx m_{\tilde\chi_1^0}\approx M_1=M_2$ which indicates the on-shell decay of $\tilde\chi_1^\pm/\tilde\chi_2^0$ to $\tilde\chi_1^0,\;W/Z$ is forbidden in this case. Then considering the lower bound of $M_1=M_2\approx m_{\tilde\chi_1^\pm/\tilde\chi_2^0}>300\;{\rm GeV}$ set by the search of $\tilde\chi_1^\pm/\tilde\chi_2^0$ off-shell decay to $\tilde\chi_1^0$ and $W/Z$, we take $m_0\equiv\mu=2M_{1,2}>600\;{\rm GeV}$ in the following analyses. In addition, the results from the LHC~\cite{Aad:2020qnn,Sirunyan:2020eab} exclude the slepton masses less than about $700\;{\rm GeV}$, we take  $M_E\gtrsim 0.8\;{\rm TeV}$ in the following analyses. Finally, for parameters in the squark sector we set $T_u=Y_u\times {\rm diag}(1,1,A_t)\;{\rm TeV}$, $T_d=Y_d\times {\rm diag}(1,1,A_b)\;{\rm TeV}$, $T_{x,e}=Y_{x,e}\times {\rm diag}(1,1,1)\;{\rm TeV}$, $m_{\tilde q}=m_{\tilde u}={\rm diag}(2,2,M_{\tilde t})\;{\rm TeV}$, $m_{\tilde d}={\rm diag}(2,2,M_{\tilde b})\;{\rm TeV}$ for simplicity.

Based on our previous analyses on the muon anomalous MDM~\cite{Yang:2018guw,hLFV,hZr,Zhang:2014iva,Yang:2020bmh,Dong:2020ioc}, the Higgs boson mass, the Higgs boson decays and $\bar B\rightarrow X_s\gamma$ in the B-LSSM, the muon anomalous MDM depends on $M_E,\mu,\tan\beta$; the Higgs boson mass depends on $A_t,\; M_{\tilde t},\;\tan\beta$; the Higgs boson decay modes $h\rightarrow gg,ZZ,WW,b\bar b,\tau\bar\tau$ depend on $A_t,\;A_b,\;\mu,\;M_{\tilde t},\;M_{\tilde b},\;\tan\beta,\;M_{H^\pm}$ (charged Higgs boson mass), the $B$ meson rare decay process $\bar B\rightarrow X_s\gamma$ depends on $M_{H^\pm},\;A_t,\;M_{\tilde t},\;\tan\beta$. In addition, the new parameters $g_{_B},\;g_{_{YB}},\;\tan\beta'$ can affect contributions to all these measured quantities. The updated experimental data on searching $Z'$ shows $M_{Z'}>4.05 \;{\rm TeV}$ with $95\%$ confidence level (CL)~\cite{171}. We take $M_{Z'}=4.2 \;{\rm TeV}$ without losing generality because the contributions from $Z'$ are suppressed by its heavy mass. The upper bound on the ratio between
the $Z'$ boson mass and its gauge coupling $g_{_B}$ at $95\%$ CL is given by~\cite{172,173}
\begin{eqnarray}
&&M_{Z'}/g_{_B}>6\;{\rm TeV}.
\end{eqnarray}
Since we focus on the muon anomalous MDM in this work, we can reasonably take $M_{H^\pm}=1.5\;{\rm TeV}$ which coincides with the experimental results of the Higgs boson mass, Higgs boson decays and $\bar B\rightarrow X_s\gamma$ well according to our previous works. And we take the Yukawa coupling constant $Y_x={\rm diag}(0.1,0.1,0.1)$ which corresponds to the Majorana mass term of right handed neutral leptons. All parameters fixed above affect the numerical results of the muon anomalous MDM and the transition magnetic moments of Majorana neutrinos negligibly. Then in order to show the allowed ranges of $\mu,\tan\beta,g_{_B},g_{_{YB}},\tan\beta'$ under the constraints from the Higgs boson mass, the Higgs boson decays and $\bar B\rightarrow X_s\gamma$, we scan the following parameter space
\begin{eqnarray}
&&M_{\tilde t}=(1.6,\;4)\;{\rm TeV},\;M_{\tilde b}=(1.6,\;4)\;{\rm TeV},\;A_{t}=(-2,\;2)\;{\rm TeV},\;A_{b}=(-2,\;2)\;{\rm TeV},\nonumber\\
&&\tan\beta=(5,\;40),\;m_0=(0.6,\;4)\;{\rm TeV},\;\tan\beta'=(1.02,\;1.5),\;g_{_B}=(0.1,\;0.7),\nonumber\\
&&g_{_{YB}}=(-0.8,\;0).\label{PS1}
\end{eqnarray}
\begin{figure}
\setlength{\unitlength}{1mm}
\centering
\includegraphics[width=3.08in]{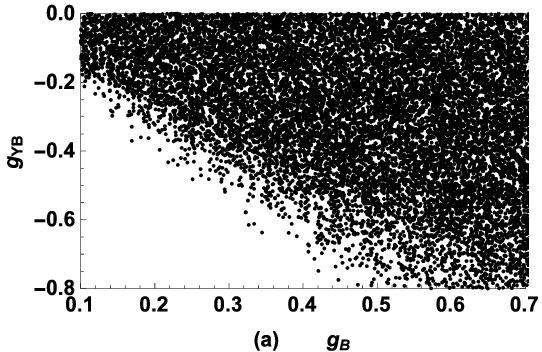}
\vspace{0.5cm}
\includegraphics[width=3in]{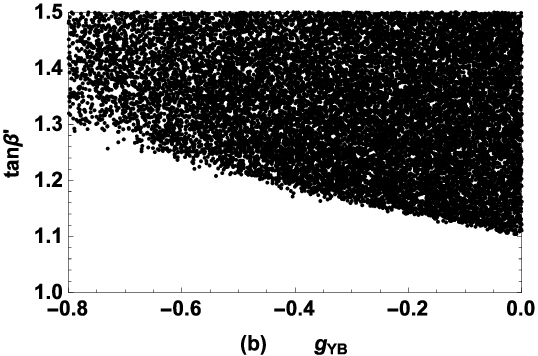}
\vspace{0cm}
\caption{Scanning the parameters in Eq. (\ref{PS1}) and considering the constraints from the observed Higgs boson mass, the Higgs boson decays, $\bar B\rightarrow X_s\gamma$ and the bound from the LHC searches on chargino/neutralino mass, we plot the allowed ranges of $\tan\beta',g_{_B},g_{_{YB}}$.}
\label{ranges}
\end{figure}
In the scanning, we keep the lightest Higgs boson mass in experimental $3\sigma$ interval, the branching ratio $Br(\bar B\rightarrow X_s\gamma)$ in experimental $3\sigma$ interval, and the Higgs boson decay modes shown in Eq. (\ref{hdecay}) in experimental $3\sigma$ interval. It can be noted that we do not scan the slepton mass parameter $M_E$ in Eq.~(\ref{PS1}), because $M_E$ affects the numerical results of $Br(\bar B\rightarrow X_s\gamma)$, Higgs boson mass and Higgs boson decay modes mentioned above negligibly, hence we only consider the constraints on the slepton masses from LHC, i.e. $M_E\gtrsim0.8\;{\rm TeV}$ in the analyses. Then we plot the allowed ranges of $g_{_B},g_{_{YB}},\tan\beta'$ in Fig.~\ref{ranges}. In our chosen parameter space, the ranges of $\mu$, $\tan\beta$ are not limited by the constraints considered above, hence we do not plot them in the figure. As shown in Fig.~\ref{ranges}, the range of $g_{_{YB}}$ is not limited by the constraints considered above when $g_{_B}\gtrsim0.4$ and $\tan\beta'\gtrsim1.3$. When $g_{_{YB}}$ approaches to $0$, $g_{_B},\tan\beta'$ are limited in the ranges $0.1\lesssim g_{_B}\lesssim0.7$, $1.1\lesssim \tan\beta'\lesssim1.5$ respectively.

\begin{figure}
\setlength{\unitlength}{1mm}
\centering
\includegraphics[width=3in]{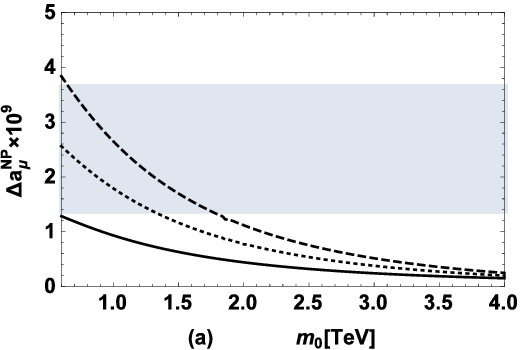}
\vspace{0.5cm}
\includegraphics[width=3in]{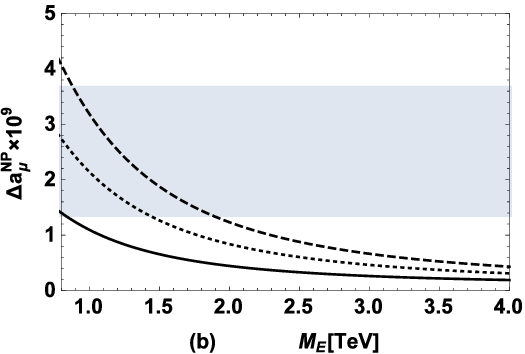}
\vspace{0cm}
\caption{$\Delta a_\mu^{NP}$ versus $m_0$ is plotted for $M_E=1\;{\rm TeV}$ (a), where the gray area denotes the experimental $2\sigma$ interval, the solid, dotted, dashed lines denote the numerical results for $\tan\beta=10,20,30$ respectively. Similarly, $\Delta a_\mu^{NP}$ versus $M_E$ is plotted for $m_0=0.8\;{\rm TeV}$ (b).}
\label{amuold}
\end{figure}
Considering the constraints shown in Fig.~\ref{ranges}, we estimate the predicted results of the muon anomalous MDM in the B-LSSM. Firstly, we study the effects of $M_E,m_0,\tan\beta$, and the new parameters in the B-LSSM are taken as $g_{_B}=0.4,g_{_{YB}}=-0.4,\tan\beta'=1.15,M_{BL}=0.6\;{\rm TeV},M_{BB'}=0.5\;{\rm TeV},\mu'=0.8\;{\rm TeV}$. Then taking $M_E=1\;{\rm TeV}$, we plot $\Delta a_\mu^{NP}$ versus $m_0$ in Fig.~\ref{amuold} (a), where the gray area denotes the experimental $2\sigma$ interval, the solid, dotted, dashed lines denote the numerical results for $\tan\beta=10,20,30$ respectively. Similarly, taking $m_0=0.8\;{\rm TeV}$, we plot $\Delta a_\mu^{NP}$ versus $M_E$ in Fig.~\ref{amuold} (b). As shown in the picture, the contributions in the B-LSSM are enhanced by large $\tan\beta$, but the effects of $\tan\beta$ are suppressed when $m_0$ is large. At the one-loop level, the dominant contributions to the muon anomalous MDM come from the chargino-sneutrino loop. With the increasing of $m_0$, the contributions from chargino-sneutrino loop are suppressed by large chargino mass which leads to that the effects of $\tan\beta$ are suppressed in this case. And the contributions are also suppressed by large $M_E$, because the masses of sleptons increase with the increasing of $M_E$. The numerical results show that the new measured muon anomalous MDM favors small $m_0$, small $M_E$ and large $\tan\beta$, i.e. $m_0\lesssim1.3 (1.7)\;{\rm TeV}$ for $\tan\beta= 20(30),\;M_E=1\;{\rm TeV}$; $M_E\lesssim1.4(1.9)\;{\rm TeV}$ for $\tan\beta=20(30),\;m_0=0.8\;{\rm TeV}$. Then in the following analyses, we take $\tan\beta=20,\;m_0=0.8\;{\rm TeV},\;M_E=1\;{\rm TeV}$ to analyze the effects of new parameters in the B-LSSM.

\begin{figure}
\setlength{\unitlength}{1mm}
\centering
\includegraphics[width=3in]{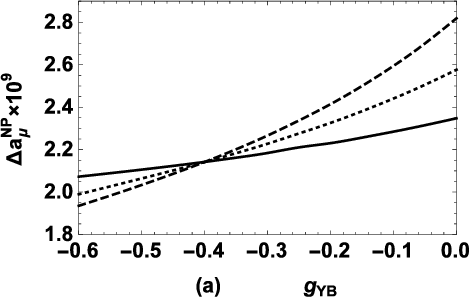}
\vspace{0.5cm}
\includegraphics[width=3in]{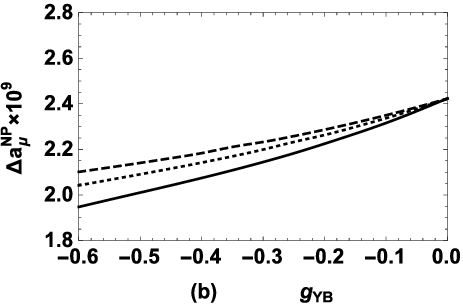}
\vspace{0cm}
\caption{Taking $M_{BL}=0.6\;{\rm TeV},M_{BB'}=0.5\;{\rm TeV},\mu'=0.8\;{\rm TeV}$, we plot $\Delta a_\mu^{NP}$ versus $g_{_{YB}}$ for $g_{_B}=0.4$ (a), where the solid, dotted, dashed lines denote the numerical results for $\tan\beta'=1.15,1.3,1.45$ respectively. Similarly, $\Delta a_\mu^{NP}$ versus $g_{_{YB}}$ is plotted for $\tan\beta'=1.2$ (b), where the solid, dotted, dashed lines denote the numerical results for $g_{_B}=0.3,0.4,0.5$ respectively.}
\label{amu-gYB}
\end{figure}
Taking the new mass terms in the neutralino sector as $M_{BL}=0.6\;{\rm TeV},\:M_{BB'}=0.5\;{\rm TeV},\:\mu'=0.8\;{\rm TeV}$, we plot $\Delta a_\mu^{NP}$ versus $g_{_{YB}}$ for $g_{_B}=0.2$ in Fig.~\ref{amu-gYB} (a), where the solid, dotted, dashed lines denote the numerical results for $\tan\beta'=1.15,1.3,1.45$ respectively. Similarly, $\Delta a_\mu^{NP}$ versus $g_{_{YB}}$ for $\tan\beta'=1.2$ is plotted in Fig.~\ref{amu-gYB} (b), where the solid, dotted, dashed lines denote the numerical results for $g_{_B}=0.3,0.4,0.5$ respectively. The picture shows $g_{_B},g_{_{YB}},\tan\beta'$ can affect the theoretical predictions of the muon anomalous MDM. It can be noted in Fig.~\ref{amu-gYB} (a) that the effects of $\tan\beta'$ are suppressed when $g_{_{YB}}=-0.4$, because $\tan\beta'$ affects the numerical results mainly through affecting the masses of slepton which depend on $\tan\beta'$ negligibly when $g_{_B}=-g_{_{YB}}$ as shown in Eq.~(\ref{ME2}). In addition, Fig.~\ref{amu-gYB} (b) shows $g_{_B}$ affects the numerical results negligibly when $g_{_{YB}}$ approaches to $0$, which indicates that $g_{_B}$ affects the numerical results mainly through the gauge kinetic mixing effect.

\begin{figure}
\setlength{\unitlength}{1mm}
\centering
\includegraphics[width=3in]{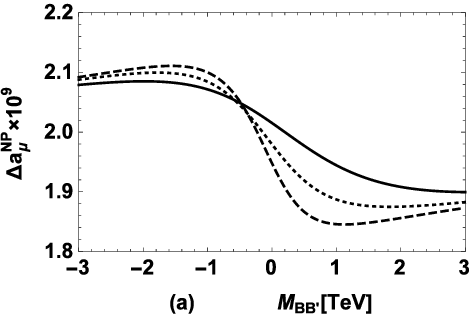}
\vspace{0.5cm}
\includegraphics[width=3in]{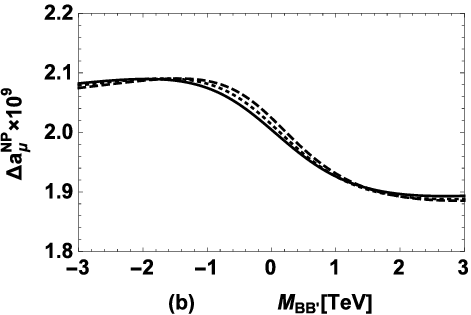}
\vspace{0cm}
\caption{Taking $g_{_{YB}}=-0.4,g_{_B}=0.2,\tan\beta'=1.2$, we plot $\Delta a_\mu^{NP}$ versus $M_{BB'}$ for $M_{BL}=0.5\;{\rm TeV}$ (a), where the solid, dotted, dashed lines denote the numerical results for $\mu'=0.5,1.5,2.5\;{\rm TeV}$ respectively. Similarly, $\Delta a_\mu^{NP}$ versus $M_{BB'}$ for $\mu'=0.8\;{\rm TeV}$ (b) is plotted, where the solid, dotted, dashed lines denote the numerical results for $M_{BL}=0.5,1.5,2.5\;{\rm TeV}$ respectively.}
\label{amu-mBB}
\end{figure}
With respect to the MSSM, there are new neutralinos in the B-LSSM, then in order to estimate the contributions from new neutralinos in the B-LSSM, we take $g_{_{YB}}=-0.4,g_{_B}=0.2,\tan\beta'=1.2$ and plot $\Delta a_\mu^{NP}$ versus $M_{BB'}$ for $M_{BL}=0.5\;{\rm TeV}$ in Fig.~\ref{amu-mBB} (a), where the solid, dotted, dashed lines denote the numerical results for $\mu'=0.5,1.5,2.5\;{\rm TeV}$ respectively. Similarly, $\Delta a_\mu^{NP}$ versus $M_{BB'}$ for $\mu'=0.8\;{\rm TeV}$ is plotted in Fig.~\ref{amu-mBB} (b), where the solid, dotted, dashed lines denote the numerical results for $M_{BL}=0.5,1.5,2.5\;{\rm TeV}$ respectively. As shown in Fig.~\ref{amu-mBB} (a), the effect of $M_{BB'}$ is affected by the value of $\mu'$ complicatedly, and the effect of $\mu'$ is highly suppressed when $|M_{BB'}|$ is large. Fig.~\ref{amu-mBB} (b) shows that $M_{BL}$ affects the numerical results negligibly. Compared with the MSSM, new neutralinos in the B-LSSM can also affect the theoretical prediction of the muon anomalous MDM.

\begin{figure}
\setlength{\unitlength}{1mm}
\centering
\includegraphics[width=3in]{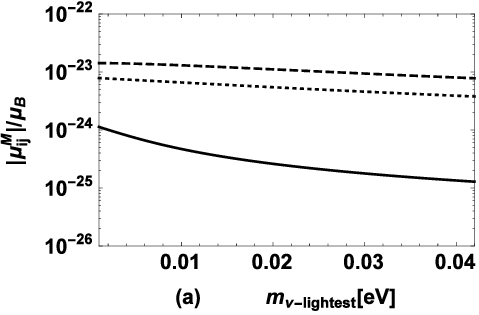}
\vspace{0.5cm}
\includegraphics[width=3in]{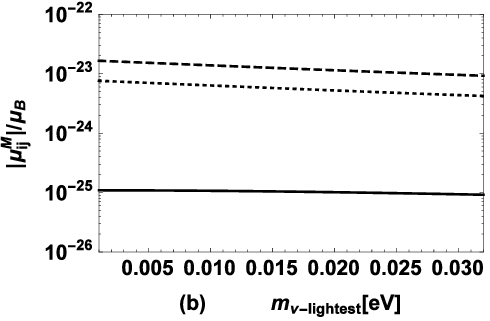}
\vspace{0cm}
\caption{$|\mu_{ij}^M|/\mu_B$ versus $m_{\nu-{\rm lightest}}$ are plotted for NH (a) and IH (b) neutrino mass, where the solid, dotted, dashed lines denotes the results of $|\mu_{12}^M|/\mu_B$, $|\mu_{13}^M|/\mu_B$, $|\mu_{23}^M|/\mu_B$ respectively.}
\label{nu-mnu}
\end{figure}
With the minimal flavor violation assumption, the contributions to the transition magnetic moments of Majorana neutrinos in the B-LSSM are dominated by the $W$-charged lepton loop shown in Fig.~\ref{nuinujr}. Hence the numerical results mainly depend on the lightest neutrino mass. Then $|\mu_{ij}^M|/\mu_B$ versus $m_{\nu-{\rm lightest}}$ are plotted in Fig.~\ref{nu-mnu} (a), (b) for NH and IH neutrino masses respectively, where the solid, dotted, dashed lines denote the results of $|\mu_{12}^M|/\mu_B$, $|\mu_{13}^M|/\mu_B$, $|\mu_{23}^M|/\mu_B$ respectively. The studies on the effect of Majorana transition magnetic moments inside the core of supernova explosions~\cite{nu47,nu48} show that the moments may leave a potentially observable imprint on the energy spectra of neutrinos and antineutrinos from supernovae even if the moments as small as $10^{-24}\mu_B$. And it can be noted from the picture that $|\mu_{13}^M|$, $|\mu_{23}^M|$ can reach $10^{-23}\mu_B$ in the B-LSSM both for NH and IH neutrino masses, which indicates that the transition magnetic moments of Majorana neutrinos predicted in the B-LSSM have great opportunity to be observed.

\section{Summary\label{sec4}}

Considering the constraints from the observed Higgs boson mass, Higgs boson decay modes $h\rightarrow \gamma\gamma,\;WW,\;ZZ,\; b\bar b,\;\tau\bar\tau$ and B meson rare decay $\bar B\rightarrow X_s\gamma$, we estimate the theoretical prediction on the muon anomalous MDM at the two-loop level in the B-LSSM. The numerical results indicate that the deviation between the SM theoretical prediction and experimental observables on the muon anomalous MDM is remedied well in the B-LSSM. Choosing the tolerance of $\Delta a_\mu^{NP}$ with $2$ standard deviations, we find that the updated muon anomalous MDM favors large $\tan\beta$ ($\tan\beta\gtrsim10$), light sleptons ($M_E\lesssim1.9\;{\rm TeV}$), light charginos and light neutralinos ($m_0\lesssim1.7\;{\rm TeV}$). In the parameter space which accommodates the muon $g-2$ anomaly, the theoretical predictions on masses of the lightest chargino, neutralino, charged sleptons and sneutrinos are well below TeV scale which are accessible at the LHC. The regions with $m_0\lesssim1.5\;{\rm TeV}$ will be probed by the next generation of dark matter direct detection experiments in the near future. Some of the CP-odd Higgs resonances with $\tan\beta\gtrsim20$ may be tested by searches of CP-even or CP-odd Higgs boson decay to $\tau^+\tau^-$ at the LHC~\cite{Ahmed:2020lua}.

In the B-LSSM, the super partners of new gauge boson and two scalar singlets mix with the neutralinos in the MSSM. The numerical results show that these new neutralinos can make contributions to the muon anomalous MDM. And new gauge coupling constants $g_{_B},\;g_{_{YB}}$ and $\tan\beta'$ (defined as the ratio of the VEVs of two new scalar singlets) in the B-LSSM can also affect the theoretical predictions of the muon anomalous MDM. In addition, the Majorana neutrinos transition magnetic moments $|\mu_{13}^M|$, $|\mu_{23}^M|$ predicted in the B-LSSM can reach $10^{-23}\mu_B$, which has opportunity to be observed on the energy spectra of neutrinos from supernovae.

\begin{acknowledgments}

The work has been supported by the National Natural Science Foundation of China (NNSFC) with Grants No. 11705045, No. 11535002, No. 12075074, the youth top-notch talent support program of the Hebei Province,  and Midwest Universities Comprehensive Strength Promotion project.

\end{acknowledgments}

\begin{appendix}
\section{The mass matrices of up squarks, down squarks and sneutrinos\label{appA}}

On the basis $(\tilde u_L,\;\tilde u_R)$, the mass matrix of up squarks is given by
\begin{eqnarray}
&&m_{\tilde u}^2=
\left(\begin{array}{cc}m_{uL},&\frac{1}{\sqrt2}(v_2 T_u^\dagger-v_1\mu Y_u^\dagger)\\\frac{1}{\sqrt2}(v_2 T_u-v_1\mu^* Y_u),&m_{uR}\end{array}\right),\label{A3}
\end{eqnarray}
with
\begin{eqnarray}
&&m_{uL}=m_{\tilde q}^2+\frac{1}{24}\Big[2g_{_B}(g_{_B}+g_{_{YB}})(u _2^2-u_1^2)+g_{_{YB}}(g_{_B}+g_{_{YB}})(v_2^2-v_1^2)\nonumber\\
&&\qquad\;\quad\;+(-3g_{_2}^2+g_{_1}^2)(v_2^2-v_1^2)\Big]+\frac{v_2^2}{2}Y_u^\dagger Y_u,\nonumber\\
&&m_{uR}=m_{\tilde u}^2+\frac{1}{24}\Big[g_{_B}(g_{_B}+4g_{_{YB}})(u _1^2-u_2^2)+g_{_{YB}}(g_{_B}+4g_{_{YB}})(v_1^2-v_2^2)\nonumber\\
&&\qquad\;\quad\;+4g_{_1}^2(v_1^2-v_2^2)\Big]+\frac{v_1^2}{2}Y_u^\dagger Y_u.\label{massu}
\end{eqnarray}

On the basis $(\tilde d_L,\;\tilde d_R)$, the mass matrix of down squarks is given by
\begin{eqnarray}
&&m_{\tilde d}^2=
\left(\begin{array}{cc}m_{dL},&\frac{1}{\sqrt2}(v_1 T_d^\dagger-v_2\mu Y_d^\dagger)\\\frac{1}{\sqrt2}(v_1 T_d-v_2\mu^* Y_d),&m_{dR}\end{array}\right),\label{A1}
\end{eqnarray}
with
\begin{eqnarray}
&&m_{dL}=\frac{1}{24}\Big(2g_{_B}(g_{_B}+g_{_{YB}})(u_2^2-u_1^2)+(3g_2^2+g_1^2+g_{_{YB}}^2+
g_{_B}g_{_{YB}})(v_2^2-v_1^2)\Big)\nonumber\\
&&\qquad\;\quad\;+m_{\tilde q}^2+\frac{v_1^2}{2}Y_d^\dagger Y_d,\nonumber\\
&&m_{dR}=\frac{1}{24}\Big(2g_{_B}(g_{_B}-2g_{_{YB}})(u_1^2-u_2^2)+2(g_1^2+g_{_{YB}}^2-
\frac{1}{2}g_{_B}g_{_{YB}})(v_2^2-v_1^2)\Big)\nonumber\\
&&\qquad\;\quad\;+m_{\tilde d}^2+\frac{v_1^2}{2}Y_d^\dagger Y_d.\label{massd}
\end{eqnarray}

Neglecting the tiny Yukawa coupling constant $Y_\nu$ (corresponding to the Dirac mass term of neutral leptons), we give the mass matrix for CP-odd sneutrinos on the basis $(\tilde \sigma_L,\;\tilde \sigma_R)$ as
\begin{eqnarray}
&&m_{\tilde \nu_{\rm odd}}^2=
\left(\begin{array}{cc}m_{\nu_{\rm odd} L},&0\\0,&m_{\nu_{\rm odd} R}\end{array}\right),
\end{eqnarray}
with,
\begin{eqnarray}
&&m_{\nu_{\rm odd} L}=\frac{1}{8}\Big(2g_{_B}(g_{_B}+g_{_{YB}})(u_1^2-u_2^2)+(g_2^2+g_1^2+g_{_{YB}}^2+
g_{_B}g_{_{YB}})(v_1^2-v_2^2)\Big)+m_{\tilde l}^2,\nonumber\\
&&m_{\nu_{\rm odd} R}=\frac{1}{8}\Big(2g_{_B}^2(u_2^2-u_1^2)+g_{_B}g_{_{YB}}(v_2^2-v_1^2)\Big)+m_{\tilde \nu}^2+\sqrt2 u_2\Re(Y_x\mu'^*)\nonumber\\
&&\qquad\;\quad\;\;\;+\sqrt2 u_1\Big(2u_1\Re(Y_xY_x^*)-\Re(T_x)\Big),
\end{eqnarray}
the mass matrix for CP-even sneutrinos is given on the basis $(\tilde \phi_L,\;\tilde \phi_R)$
\begin{eqnarray}
&&m_{\tilde \nu_{\rm even}}^2=
\left(\begin{array}{cc}m_{\nu_{\rm even} L},&0\\0,&m_{\nu_{\rm even} R}\end{array}\right),
\end{eqnarray}
with
\begin{eqnarray}
&&m_{\nu_{\rm even} L}=\frac{1}{8}\Big(2g_{_B}(g_{_B}+g_{_{YB}})(u_1^2-u_2^2)+(g_2^2+g_1^2+g_{_{YB}}^2+
g_{_B}g_{_{YB}})(v_1^2-v_2^2)\Big)+m_{\tilde l}^2,\nonumber\\
&&m_{\nu_{\rm even} R}=\frac{1}{8}\Big(2g_{_B}^2(u_2^2-u_1^2)+g_{_B}g_{_{YB}}(v_2^2-v_1^2)\Big)+m_{\tilde \nu}^2-\sqrt2 u_2\Re(Y_x\mu'^*)\nonumber\\
&&\qquad\;\quad\;\;\;+\sqrt2 u_1\Big(2u_1\Re(Y_xY_x^*)+\Re(T_x)\Big),
\end{eqnarray}
where $\tilde \nu_{L,R}\equiv \tilde\phi_{L,R}+i\tilde\sigma_{L,R}$.

\end{appendix}

\end{document}